\RequirePackage{fix-cm}
\RequirePackage{amsmath}

\documentclass[twocolumn,epjc3]{svjour3}  
\smartqed  
\RequirePackage{graphicx}
\RequirePackage{lineno}
\usepackage[T1]{fontenc}          
\usepackage{url}                  
\usepackage[hidelinks]{hyperref}  
\journalname{Eur. Phys. J. C}
\begin{document}

\title{Leakage current evolution in LHCb VELO sensors during the LHC Run 1 and Run 2 data-taking periods
}


\author{M. Pycior\thanksref{e1,a1}
\and
T. Szumlak\thanksref{e2,a1}
\and
A. Oblakowska-Mucha\thanksref{a1}
\and
K. Akiba\thanksref{a2}
\and
W. Barter\thanksref{a3}
\and
S. Borghi\thanksref{a4}
\and
T. Bowcock\thanksref{a5}
\and
E. Buchanan\thanksref{a3}
\and
J. Buytaert\thanksref{a6}
\and
S. de Capua\thanksref{a4}
\and
S. Chen\thanksref{a7}
\and
P. Collins\thanksref{a6}
\and
F. Dettori\thanksref{a8}
\and
L. Eklund\thanksref{a9}
\and
T. Evans\thanksref{a2}
\and
M. Gersabeck\thanksref{a10}
\and
T. Gershon\thanksref{a11}
\and
T. Hadavizadeh\thanksref{a12}
\and
K. Hennessy\thanksref{a5}
\and
W. Hulsbergen\thanksref{a2}
\and
D. Hutchcroft\thanksref{a5}
\and
M. John\thanksref{a13}
\and
P. Kopciewicz\thanksref{a6}
\and
P. Koppenburg\thanksref{a2}
\and
T. Latham\thanksref{a11}
\and
M. Majewski\thanksref{a1}
\and
C. Parkes\thanksref{a4}
\and
A. Poluektov\thanksref{a14}
\and
W. Qian\thanksref{a15}
\and
K. Rinnert\thanksref{a5}
\and
E. Rodrigues\thanksref{a5}
\and
M. Schiller\thanksref{a16}
\and
M. Smith\thanksref{a17}
\and
M. van Beuzekom\thanksref{a2}
\and
J. Velthuis\thanksref{a18}
\and
M. Williams\thanksref{a3}
}

\thankstext{e1}{e-mail: mpycior@agh.edu.pl, first author}
\thankstext{e2}{e-mail: szumlak@agh.edu.pl, corresponding author}


\institute{Faculty of Physics and Applied Computer Science, AGH University of Krakow, Krakow, Poland\label{a1}
\and
Nikhef National Institute for Subatomic Physics, Amsterdam, The Netherlands\label{a2}
\and
School of Physics and Astronomy, College of Science and Engineering, Edinburgh,
United Kingdom\label{a3}
\and
School of Physics and Astronomy, University of Manchester, Manchester, United Kingdom\label{a4}
\and
University of Liverpool, Liverpool, United Kingdom\label{a5}
\and
European Organization for Nuclear Research (CERN), Geneva, Switzerland\label{a6}
\and
Institute of High Energy Physics (IHEP), Chinese Academy of Sciences, China\label{a7}
\and
Universita degli Studi di Cagliari, Universita e INFN, Cagliari, Italy\label{a8} 
\and
Department of Physics and Astronomy, Uppsala Universitet, Sweden\label{a9} 
\and
Physikalisches Institut, Universitat Freiburg, Germany\label{a10} 
\and
Department of Physics, University of Warwick, United Kingdom\label{a11} 
\and
School of Physics and Astronomy Monash University, Monash University, Australia\label{a12}
\and
Department of Physics, University of Oxford, Oxford, United Kingdom\label{a13}
\and
Aix Marseille Univ, CNRS/IN2P3, CPPM, Marseille, France\label{a14} 
\and
University of Chinese Academy of Sciences, China\label{a15} 
\and
Physikalisches Institut, Ruprecht Karls Universitaet Heidelberg, Germany\label{a16}
\and
Imperial College, London, United Kingdom\label{a17} 
\and
School of Physics, University of Bristol, United Kingdom\label{a18} 
}

\date{Received: date / Accepted: date}

\maketitle

\begin{abstract}
Silicon detectors operating in the harsh mixed-hadron radiation environment of modern high-energy physics experiments suffer progressive bulk damage, manifesting macroscopically as a steady rise in leakage current. The Vertex Locator (VELO) of the LHCb experiment provides a unique opportunity to study this phenomenon: positioned only millimeters from the colliding LHC proton beams, its sensors are the silicon devices closest to any LHC interaction point and were arguably the most heavily irradiated silicon detectors in operation at the LHC throughout Run 1 and Run 2 (2011–2018), accumulating fluence of up to $10^{15}$ n$_\mathrm{eq}$/cm$^2$.
The eight years of continuous detector monitoring data collected over this period form a uniquely comprehensive dataset, enabling a high-precision comparison of radiation-damage models with extreme mixed-hadronic radiation conditions that no other silicon system at the LHC can match. In this work, the Hamburg model is applied to predict the evolution of the leakage current in VELO sensors, and its predictions are compared in detail with the measurements. Preliminary studies addressing temperature measurement uncertainties yield results consistent with the model's underlying assumptions. A weaker-than-expected dependence of the leakage current on sensor radial position is observed, an effect whose likely origin is traced to the fluence simulation rather than to the damage model itself.
\end{abstract}
\section{Introduction}
\label{intro}
The Large Hadron Collider beauty (LHCb) experiment operating at the LHC is dedicated to the study of heavy-flavour physics.

During the LHC Run 1 (2011–2012) and Run 2 (2016–2018) data-taking periods, more than 9 fb$^{-1}$ of integrated luminosity was delivered to LHCb at proton–proton centre-of-mass energies ranging from 7 to 13~TeV. Since 2015, LHCb has participated in lead-proton and lead-lead runs. 

\begin{figure}
\resizebox{0.495\textwidth}{!}{%
  \includegraphics{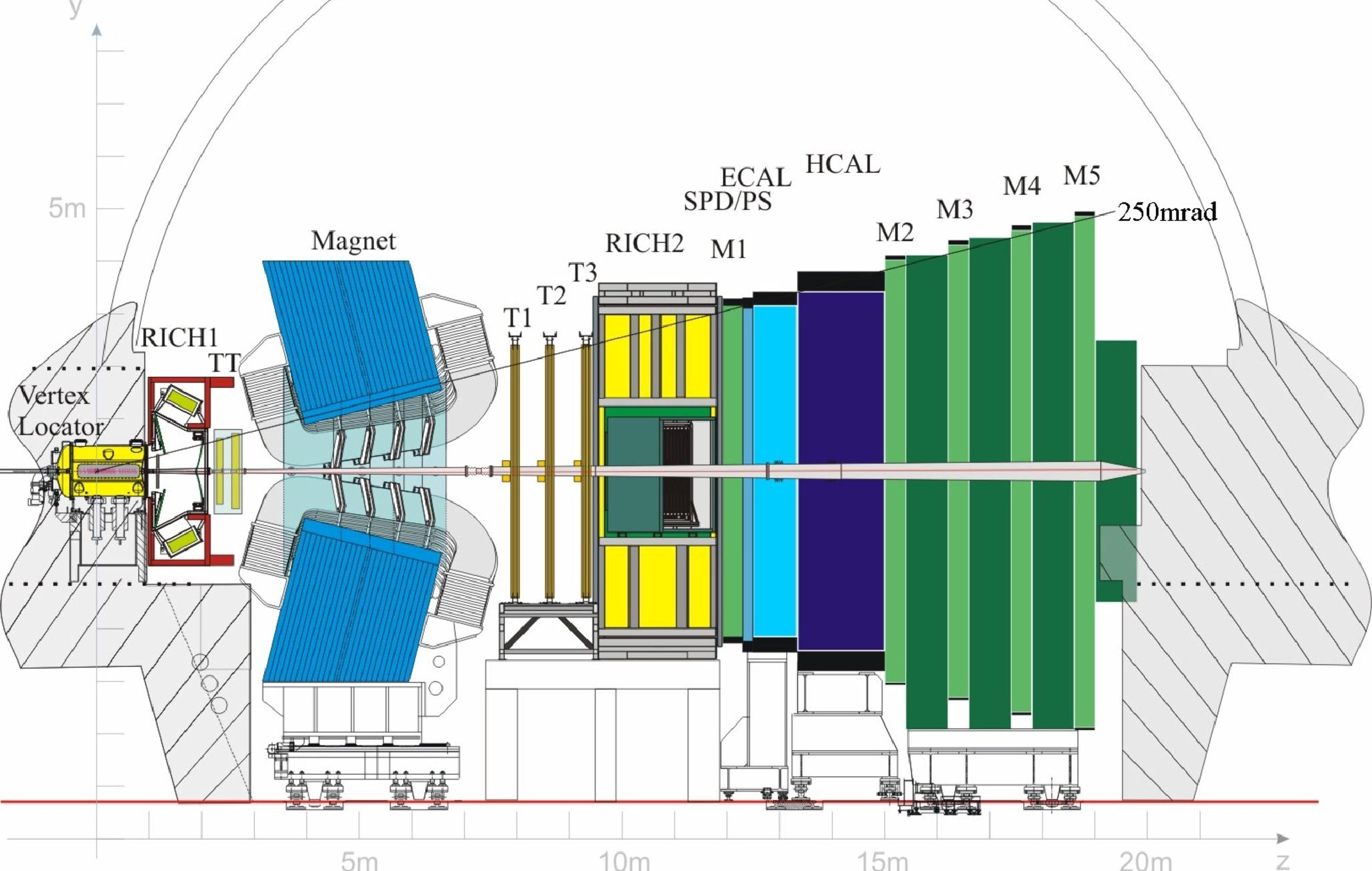}
}
\caption{Layout of the LHCb spectrometer. Vertex Locator (VELO) surrounds \textit{pp} interaction region, which is on the left side of the picture.}
\label{fig:lhcb}    
\end{figure}

The LHCb detector is a single-arm spectrometer covering the pseudorapidity range $2<\eta<5$. It includes a high-precision tracking system consisting of a silicon-strip vertex detector, the VELO, surrounding the proton-proton (\textit{pp}) interaction region, a large-area silicon-strip detector, the Trigger Tracker (TT), located upstream of a dipole magnet with a bending power of about 4 Tm; and three stations of silicon-strip detectors (Inner Tracker) and straw drift tubes (Outer Tracker) placed downstream of the magnet, referred to as the T stations, see Fig.~\ref{fig:lhcb} \cite{LHCb:2008vvz}. 

The Vertex Locator (VELO) subsystem is a silicon microstrip detector designed to reconstruct primary and secondary vertices of particles produced in \textit{pp} collisions and the first stage of the tracking system. It collected data during Runs 1 and 2 and was replaced by a new pixel design, which has been in operation since 2022 (Run~3)~\cite{Collaboration:1624070}.

The VELO modules are positioned along the beamline, aligned with the $z$-axis in the detector coordinate system. To obtain the precision vertexing required for heavy-flavour physics, the closest active silicon sensor region is located 8.2 mm
from the beam axis, while the silicon edge is located at a distance of 7 mm. This close proximity to the interaction point introduces numerous technological and design challenges, primarily due to the risk of radiation damage from the high particle density. To protect the sensors during beam injection and collision preparation phases, the detector is divided into two movable halves, which are retracted to a safe distance of 29 mm from the proton beam and are only closed for data taking, with LHC stable beam declaration. Assuming the retraction mechanism to operate reliably, and considering the strong radial dependence of the particle fluence, possible radiation damage accumulated during beam preparation was not included in the irradiation simulation presented in this paper, as its potential contribution was expected to be well below the achieved precision.

The VELO detector was designed to operate at extremely high fluence – up to 10$\times 10^{13}$  MeV n$_{\text{eq}}$cm$^{-2}$  per 1~fb$^{-1}$. The total radiation dose from ion beams is comparable to that resulting from 25~nb$^{-1}$ of $pp$ interactions; therefore this contribution is not significant for the results and was not taken into account in this analysis~\cite{Akiba:2018rhn}. 

As a simple quantitative reference, an average increase of the sensor
leakage current of approximately 18~$\mu$A per fb$^{-1}$ of delivered
luminosity was reported for the VELO in Ref.~[5]. This value was obtained
from 2010--2011 data ($\sqrt{s}=7$~TeV, about 1.2~fb$^{-1}$) for sensors
operated at the nominal bias voltage of 150~V and a mean operational
temperature of approximately $-7\,^{\circ}$C. It refers to the current
drawn by the whole sensor, averaged over all sensors excluding those
dominated by surface current, and includes the effect of several short
annealing periods. Over the full 2011--2018 dataset analysed in this paper,
the corresponding average rate is approximately 40~$\mu$A per fb$^{-1}$
when the currents are normalised to 0~$^{\circ}$C, including the
temperature correction described in Sect.~5.1 (see Fig.~9). The higher
value arises mainly from the different reference temperature, with an
additional contribution from the larger fluence per unit luminosity at
$\sqrt{s}=13$~TeV in Run~2. Since the rate depends on the centre-of-mass
energy, the sensor position, the operating temperature and the annealing
history, both values should be regarded as indicative only.



Although the sensors were originally planned to be replaced after Run~1, the
lower-than-nominal collision energy allowed them to remain in operation until
the end of Run~2. Nevertheless, radiation-damage effects became increasingly
apparent in the detector response, making their measurement and understanding
essential for correct and safe operation. At the start of operation, the
sensors had an average noise of 1.91~ADC counts, a signal-to-noise ratio of
approximately 20 and an average cluster-finding efficiency of
99.45\%~\cite{VELOperf}. Despite the growth of the leakage current, the noise
in 2018 remained at about 2~ADC counts or less for all strips, with most
probable signal values between 28 and 35~ADC counts for all
sensors~\cite{Hutchcroft2018}, corresponding to a signal-to-noise ratio of
roughly 15 or higher. The main operational consequence of the irradiation was
the increase of the effective depletion voltage following type inversion of
the innermost sensor regions. The bias voltage, initially 150~V for all
sensors, was therefore adjusted sensor by sensor on the basis of regular
charge-collection scans and reached up to 400~V. In 2018, a controlled
annealing at 20~$^{\circ}$C for 40~hours reduced the average sensor current
at 150~V from 190 to 150~$\mu$A, consistent with the Hamburg-model
prediction, and ensured that no sensor required more than 450~V during the
final year of operation, below the 500~V hardware limit~\cite{Hutchcroft2018}.
In addition, a localised loss of cluster-finding efficiency in the R sensors,
related to charge pick-up by the second-metal-layer routing lines, appeared
after the first year of operation~\cite{Hutchcroft2018}; its effect on
tracking was small and was accounted for in the simulation. Overall, the
tracking efficiency and resolution remained as expected from the design until
the end of Run~2~\cite{Hutchcroft2018}.

Monitoring of the performance of the VELO sensors throughout the eight years of LHCb data taking produced an extensive dataset of detector parameters, such as temperature, leakage current and applied bias voltage. These measurements can be used to investigate the complex problem of radiation damage modelling in silicon detectors and allow for the validation of the model commonly used for the study of the impact of particle radiation and the microscopic changes in the silicon structure on the macroscopic measurements. Such long-term data collection, combined with the unprecedentedly high radiation environment provided by the LHC, is unmatched by any other experimental facility.

The Hamburg model \cite{Moll:1999kv} for the leakage current growth in the VELO silicon sensor was implemented in dedicated simulation software. Recorded sensor temperatures and simulated fluence \cite{Akiba:2018rhn} were used to predict the evolution of the leakage current, which were subsequently validated against actual measurements. Thanks to the VELO geometry, it was also possible to study the dependence of the results on the relative position with respect to the IP and introduce the temperature correction.
\section{VELO and its radiation environment}
\label{sec:2}
\begin{figure*}
\centering
\resizebox{0.9\textwidth}{!}{%
  \includegraphics{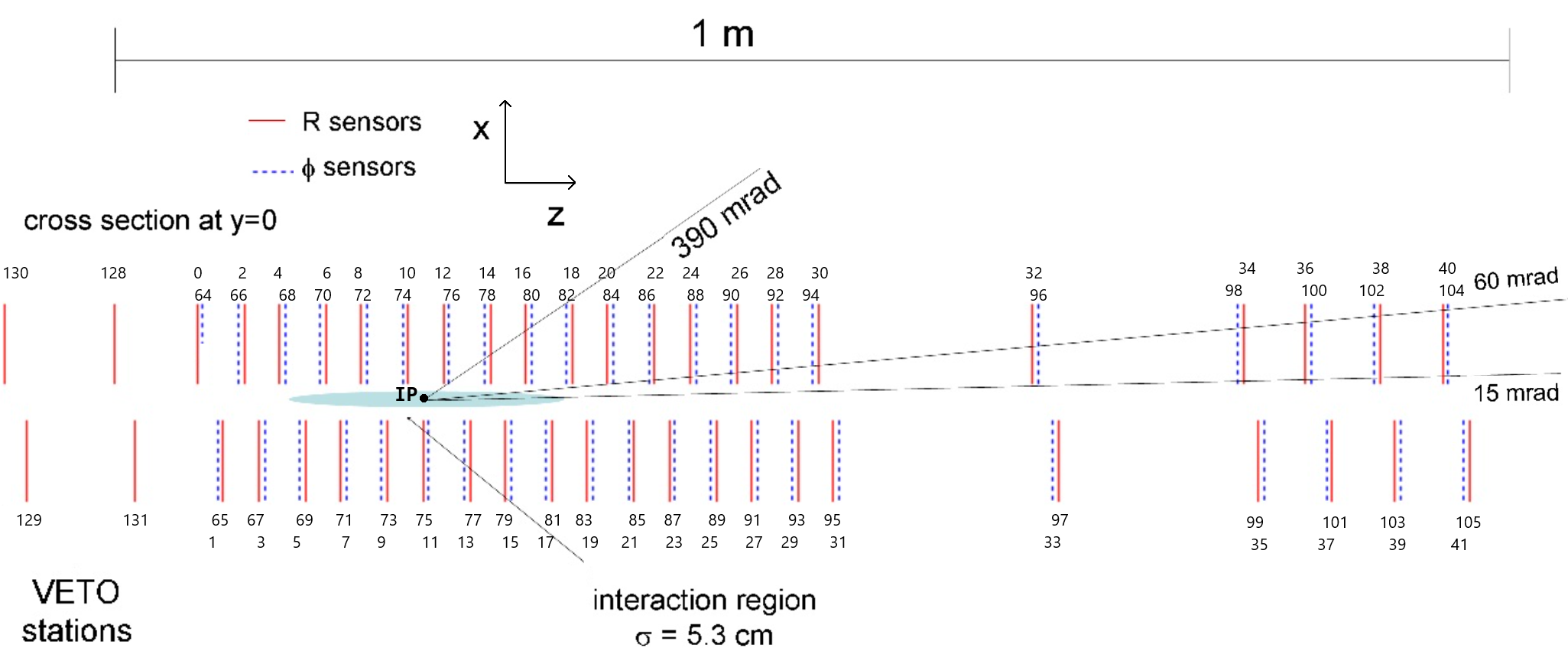}
}
\caption{Layout of the numbered VELO silicon sensors in the (x,z) plane (IP at z=0). Colours represent different sensor types. }
\label{fig:sensors_pos}    
\end{figure*}
\begin{figure}
\centering
\resizebox{0.45\textwidth}{!}{%
  \includegraphics{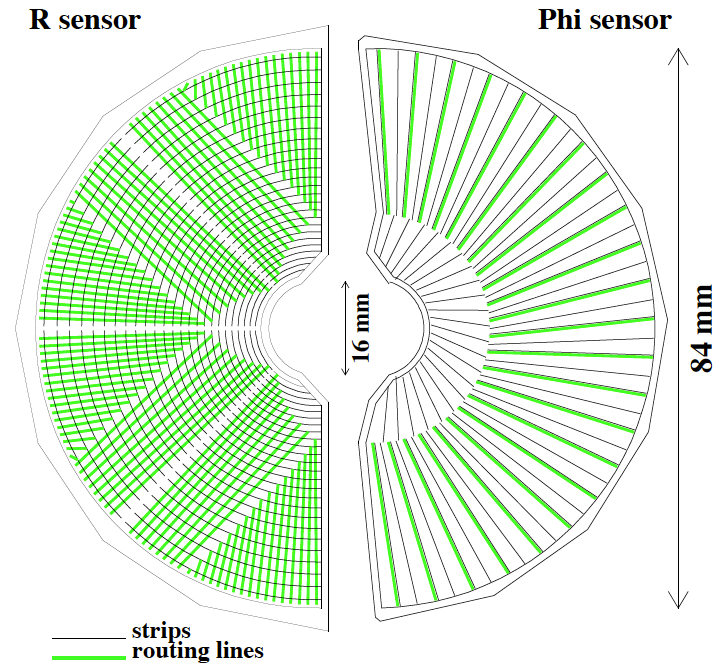}
}
\caption{Schematic representation of the strips and routing lines of an R-type and a $\Phi$-type sensor.
}
\label{fig:velo_strips}    
\end{figure}

The detector's 88 microstrip silicon sensors, mounted on modules positioned perpendicular to the beam axis, had a half-disc geometry with a diameter of 84 mm and a thickness of 300~$\mu$m, featuring a central recess to accommodate the beam. Each half of the detector contains 21 double-sided and 2 one-sided modules~-~see Fig.~\ref{fig:sensors_pos}. 
The spatial orientation of the microstrips defines the sensor type: in double-sided modules, one sensor has strips perpendicular to the radius and measures the radial coordinate, while the other has radially oriented strips and measures the azimuthal angle $\Phi$. The one-sided, upstream modules are R-type only, see Fig.~\ref{fig:velo_strips}.

The modules are positioned from $-300$ to $750$~mm along the beam axis ($z$ position), surrounding the IP, and the radius of the active volume ranges from $8.2$ to $42$~mm. This close proximity of the inner edge greatly increases exposure to radiation damage. The detector was kept in a vacuum, which is separated from the LHC vacuum by a $300~\mu$m thick aluminium foil~\cite{LHCb:2008vvz}.

\begin{figure}
\centering
\resizebox{0.45\textwidth}{!}{%
  \includegraphics{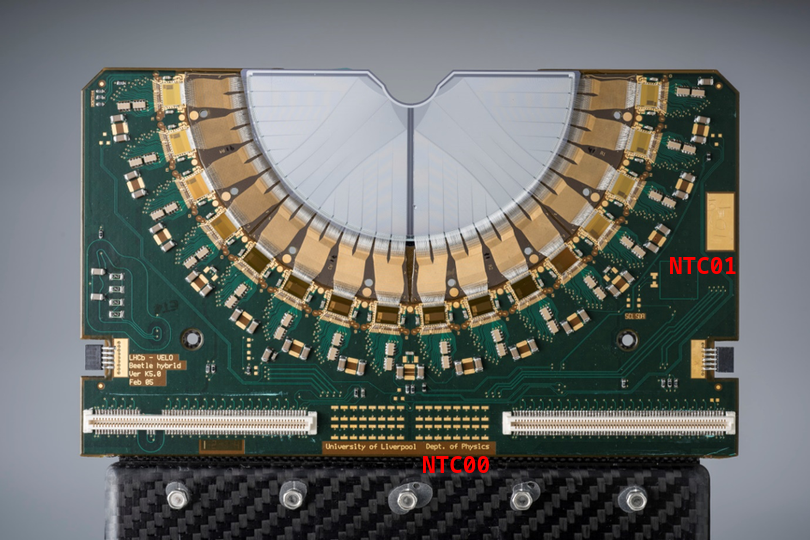}
}
\caption{The vertex locator module of microstrip VELO (one half). The silver parts in the middle of the half-circles are silicon sensors, surrounded by the Beetle readout electronics. The two gold rectangles indicate the NTC sensors: NTC01 on the right, and NTC00 next to the cooling device (on the bottom). 
}
\label{fig:velo_module}    
\end{figure}

The VELO sensors were cooled by the CO$_2$ cooling system 
to minimise the risk of thermal runaway, to reduce the effects of irradiation, and to avoid non-beneficial annealing. The operational temperature was approximately $-8$~\textdegree{}C, varying between sensors depending strongly on their $z$ positions~\cite{Affolder:2013zoa}. The detector was also kept cool during breaks from data taking; in principle, only maintenance work, temperature scans, tests, or breakdowns could cause the sensors to heat above $-5$~\textdegree{}C.
In addition to the silicon strips, cooling blocks, and front-end electronics, Negative Temperature Coefficient (NTC) sensors were located on both sides of each hybrid, as shown in Fig.~\ref{fig:velo_module}. 
The silicon sensors were biased with high voltage. Initially, the voltage required to fully deplete the active volume of the sensor was 150~V, and it increased over time due to detector ageing to 350~V by the end of 2018.

For centre-of-mass energies of several TeV, approximately 25\% of $pp$ collisions are elastic. Particles produced in inelastic interactions are responsible for the creation of most defects in the crystal lattice. The inelastic cross-section for $pp$ interaction is $\sigma_{in} = 68.7 \pm 2.1 \pm 4.5$~mb at $\sqrt{s} = 7$~TeV and $\sigma_{in} = 75.4 \pm 3.0 \pm 4.5$~mb at $\sqrt{s} = 13$~TeV, according to a 2018 study~\cite{LHCb:2018ehw}. During proton collisions, more than a hundred charged and neutral particles are produced, which pass through the detector and interact with its material. The main source of defect-creating interactions in the vertex locator region is the prompt production of hadrons, among which heavy particles, such as protons, neutrons, pions, and kaons, significantly affect the silicon lattice. 

Because of its physics program, LHCb is optimised to operate at luminosities from one to two orders of magnitude lower than the nominal LHC luminosity delivered to CMS and ATLAS. Nevertheless, owing to the exceptionally close proximity of the silicon modules to the beam, VELO remained the most irradiated detector at the LHC. By the end of Run~2, LHCb had accumulated more than 9~fb$^{-1}$ of integrated luminosity, and the estimated irradiation of the innermost parts of the sensors was as high as $6.5 \times 10^{14}$~n$_{\text{eq}}$cm$^{-2}$~\cite{DeCosa:2021nzr}.
Although numerous dosimeters are deployed throughout the cavern, direct measurement of radiation close to the interaction point is not feasible and must be supplemented by simulations. 
\section{Modelling the increase of leakage current from radiation damage}
\label{sec:1}
During proton collisions, a large number of charged and neutral particles are produced, passing through the detector and interacting with its material by ionisation or by non-ionising energy loss (NIEL), which might significantly affect the crystal structure. 

Silicon lattice defects create additional energy levels within the band-gap region, which introduce changes in the macroscopic properties of the detector by altering the electronic characteristics of the semiconductor: leakage current growth, a change in the effective doping concentration, and a reduction in charge collection efficiency due to charge carrier trapping. This effectively determines the detector’s usable lifetime. The effects described above cannot readily be reversed, and the detector cannot be restored to its initial state. However, the mobility of defects may mitigate part of the damage~\cite{Moll:1999kv}.

The leakage current in a sensor is the electrical current flowing in the absence of a particle signal. In a reverse-biased diode, it consists of two components: bulk-generation current, originating in the depleted silicon volume, and surface-generation current, arising primarily from generation centres associated with defects and interface states at the Si--SiO$_2$ boundary. This study focuses on the bulk component, which increases with irradiation and becomes dominant at high fluences, whereas the surface component tends to saturate. The increasing number of radiation-induced defects in silicon creates new generation and recombination centres. During operation, higher reverse current worsens the signal-to-noise ratio, which challenges trigger systems and particle identification software. It is highly dependent on the temperature:
\begin{equation}\label{eq:prop-current}
    I(T)\propto T^2 e^{-\frac{E_{\text{eff}}}{2k_b T}}
\end{equation}
where $k_b$ is the Boltzmann constant and $E_{\text{eff}}$ is the effective energy gap of silicon. As module power consumption increases, it heats up, creating a feedback loop that further increases the generated current. This, under extreme conditions, can lead to thermal runaway~\cite{DeCosa:2021nzr}. The leakage current increase, $\Delta I$ is expected to be proportional to depleted volume $V$  and neutron-equivalent fluence $\Phi_{eq}$, with the proportionality factor given by the current-related damage coefficient $\alpha$:
\begin{equation}\label{eq:lin_fluence}
    \Delta I = \alpha\: V\: \Phi_{\text{eq}}
\end{equation}
Leakage current can be measured precisely and continuously, making it a useful indicator of radiation damage and the received fluence, allowing comparison with predictions via $\alpha$. In contrast, the overall trend is an increase in current, $\alpha$ evolves between irradiation periods due to annealing, which has to be reliably modelled.

The most widely used approach for predicting the increase in leakage current is the Hamburg model~\cite{Moll:1999kv}. The evolution of $\alpha$, accounting for both short- and long-term annealing, is parametrised empirically as a function of the annealing time $t$ and temperature $T_a$. To account for the different time and temperature dependencies of beneficial and reverse annealing, the factors $\alpha_I$ and $\beta$ were introduced, respectively. The factor $\alpha_0^*$ describes the stable damage component.
\begin{equation} \label{eq:main_no_sums}
    \Delta I=V \cdot \Phi_{eq} \cdot\left[\alpha_{\mathrm{I}} \exp \left(- \frac{t}{\tau_I}\right)+\alpha_0^*-\beta \log \left(\ \frac{\Theta \cdot t}{t_0}\right)\right]
\end{equation}
\begin{table}
\caption{Initial parameters used in the simulation of leakage current growth according to the Hamburg model (Equation~\ref{eq:main_no_sums}).
They are valid only for the stated reference temperature~\cite{Moll:1999kv}. 
}
\label{tab:1}       
\begin{tabular*}{0.49\textwidth}{@{\extracolsep{\fill}}lll}
\hline\noalign{\smallskip}
Parameter & Value & Unit \\
\noalign{\smallskip}\hline\noalign{\smallskip}
$T_\mathrm{ref}$ & $294.15$ & K \\
$E_I^*$ & $1.30$ & eV \\
$\alpha_I$ & $1.23 \times 10^{-17}$ & A/cm \\
$\alpha_0^*$ & $7.07 \times 10^{-17}$ & A/cm \\
$\beta$ & $3.29 \times 10^{-18}$ & A/cm \\
$E_I$ & $1.11$ & eV \\
$K_{0I}$ & $1.2 \times 10^{13}$ & s$^{-1}$ \\
$t_0$ & $1.0$ & min \\
$E_\mathrm{eff}$ & $1.21$ & eV \\
\noalign{\smallskip}\hline
\end{tabular*}
\end{table}
The model parameters (see Table~\ref{tab:1}) were initially determined experimentally, before prolonged cooling became widespread, using material irradiated with a single type of source (neutron) and annealed for periods limited to timescales achievable in the laboratory (up to 1 year). To extend the model's applicability to different annealing temperatures and longer periods, additional measurements were performed under a wider range of annealing conditions. This allowed the author of the Hamburg model to parametrise the time and temperature dependence of the annealing factors and introduce the scaling functions $\tau_I$ and $\Theta$, enabling the model to describe annealing over periods of several years. The time-scaling factor $\tau_I$ characterises the rate at which defects disappear and is described by the Arrhenius relation:
\begin{equation}\label{eq:tau}
\frac{1}{\tau_I} = K_{0I} \cdot \exp\left(-\frac{E_I}{k_B T_a}\right)
\end{equation}
where $K_{0I}$ is the frequency factor, and $E_I$ is the activation energy for the process. The $\Theta$ factor scale $\beta$ to match annealing time $t$:
\begin{equation}\label{eq:big_theta}
    \Theta\left(T_a\right)=\exp \left(-\frac{E_I^*}{k_B}\left(\frac{1}{T_a}-\frac{1}{T_{\text{ref}}}\right)\right)
\end{equation}
The resulting $\Delta I$ is applicable for $T_{\text{ref}}$. As simulations have to be made for $T_{\text{ref}}=$~21~\textdegree C (the lowest temperature for which the model was parametrised), the sensor currents have to be normalised to a reference temperature, or conversely, the simulation has to be adjusted to match the collected data using the following scaling (based on the relation given in Equation~\ref{eq:prop-current}):
\begin{equation}\label{eq:R_for_i}
\frac{I(T_{\text{ref}})}{I(T)} = \left( \frac{T_{\text{ref}}}{T} \right)^2  
\exp \left[ - \frac{E_{\text{eff}}}{2 k_B} \left( \frac{1}{T_{\text{ref}}} - \frac{1}{T} \right) \right]
\end{equation}
The uncertainties associated with the original parametrisation are not reported, and its applicability at very low temperatures is uncertain, as the fit was based on measurements at only five, relatively high annealing temperatures (21, 49, 60, 80 and 106~\textdegree C). Any resulting effect on our results is therefore difficult to quantify and disentangle from the other sources of uncertainty discussed in this paper. Further studies involving analyses similar to those presented here, supported by larger datasets and a better understanding of the various factors affecting the results, could help assess this effect more reliably. 

For the original formulation of the Hamburg model and a detailed description of its parametrisation, see~\cite{Moll:1999kv}.
%
\section{Data}
The VELO sensors were continuously monitored. In addition, periodic current-temperature (IT)~\cite{Hickling:2011mxa} and current-voltage (IV)~\cite{Gureja:2011nxa} scans were performed. These scans provide high-precision measurements and valuable information about the current state of the detector, but they do not supply data with the time continuity required to accurately model annealing behaviour. IV scans are typically performed once a week, whereas IT scans are carried out a few times a year, as they can only be performed during LHC shutdowns. During the first IT scans in 2011 the maximum temperature exceeded 10~\textdegree{}C, but by Run~2 the data points were taken up to around -5~\textdegree{}C, due to the increase in leakage current.

In this analysis, an extensive experimental database containing each sensor’s temperature, leakage current, and applied bias voltage recorded over the lifetime of the VELO was used, both as a source of actual $I_{\text{leak}}$ measurements and as input for the simulation. The dataset contained numerous anomalies and an extremely large number of entries;
measurements were recorded almost continuously, including periods of beam preparation, maintenance periods, and malfunctions, which further obscured the desired trends. Therefore, dedicated software was developed for binning, filtering and visualising content. 

Before filtering, the data were binned, as the database contains up to several hundred records per second, making direct analysis of raw data impossible. Time bins of 1 and 10 minutes were used, with no noticeable impact on the results, as both are much shorter than the timescale of significant leakage current changes, typically a week or longer.

The applied voltage data were first used to filter the leakage current measurements, as only measurements taken during stable detector operation, when the sensors were fully depleted and the applied high voltage was constant, are suitable for identifying long-term evolution trends. However, this approach alone proved insufficient, as it did not remove several types of anomalies in the current evolution. These included planned operations, such as IT scans (which are performed with fully depleted sensors), as well as unplanned events, such as faulty readings. Therefore, the same approach was also applied directly to the current measurements. In this case, periods during which the current varied by more than a predefined tolerance were removed. This procedure was supported by run timestamps from the LHCb operation database and allowed us to obtain the adequately filtered leakage current measurements required for this analysis. All results were sanity-checked. Although the dataset was substantially reduced, approximately 150,000 current records per sensor were retained over the period of several years, allowing the leakage current evolution to be precisely reproduced.

The temperature data were not filtered, as they are used in the simulation and affect the annealing process. We would also like to emphasise that the filtering procedure has no connection to the leakage current simulation itself. Furthermore, the filtered measured current evolution was found to be largely insensitive to changes in the filtering parameters.

Simulating the increase in leakage current requires knowledge of each sensor's irradiation history, expressed in terms of fluence and obtained from simulations, as well as its temperature history, obtained from measurements. To determine the radiation field to which the detector was exposed, a data sample simulated in FLUKA \cite{Bohlen:2014buj} was used to obtain neutron-equivalent fluence $\Phi_{\text{eq}}$ per luminosity unit, for every sensor independently. In the FLUKA input, information about the source, geometry, materials, and magnetic field is implemented, allowing reliable scoring to be performed in selected regions. Since the leakage current is measured for the entire sensor, the average value of $\Phi_{\text{eq}}$ over the active area of each sensor was used in this study. It is worth mentioning, however, that the particle flux is highly non-uniform spatially, and fluence estimates in different radial regions of the sensors have been studied in the past (see~\cite{Akiba:2018rhn}). The fluence dataset was combined with the delivered luminosity from the LHCb database and \textit{pp} cross-section described in Section~\ref{sec:2}.

As leakage current contributions are subject to time-dependent annealing, predicted $I_{\mathrm{leak}}$ at a given time is obtained as the sum of simulated contributions $\Delta I$, each corrected for all annealing periods that occurred between its creation and the evaluation time. Temperature and fluence data, for each sensor individually, were time-binned to match the chosen 6-hour simulation step. The dependence of the results on the bin length was extensively tested using time intervals ranging from minutes to weeks, and all time steps shorter than 24 hours showed no appreciable effect on the outcome. 
\section{Results}
\begin{figure}
\centering
\resizebox{0.49\textwidth}{!}{%
  \includegraphics{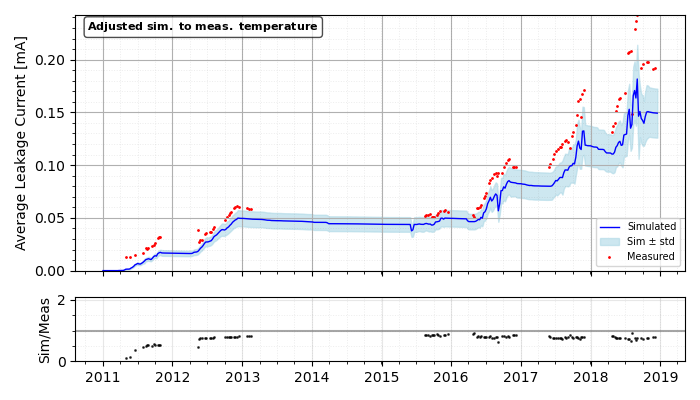}
}
\caption{Evolution of simulated and measured average $I_{\text{leak}}$. The bottom plot shows the ratio between the simulation and measurements. }
\label{fig:mean_all}
\end{figure}

The time evolutions of $I_{\text{leak}}$ were simulated for each sensor individually, and compared with the measured values. For this purpose, simulation was normalised to the NTC01 temperature using Equation~\ref{eq:R_for_i}, see Fig.~\ref{fig:mean_all}.  While the overall shape is reproduced very well, a discrepancy is visible at the beginning of the operation period (2011–2012). This can be attributed to the contribution of the surface current, which gradually disappears with irradiation (see Section~\ref{sec:across_sens}). The simulations underestimate the measured values, with a mean percentage difference of $-21.20 \%\pm5.95$ percentage points (the period dominated by the surface-current contribution was excluded from the calculation). The evolution of the leakage current reveals periods without collisions, during which annealing of the damage factor $\alpha$ occurs, effectively reducing leakage current and causing observable drops in the current profile. This is particularly evident between 2013 and 2015 (LS1) and during winter breaks. 
\subsection{Temperature correction}
\begin{figure}
\centering
\resizebox{0.49\textwidth}{!}{%
  \includegraphics{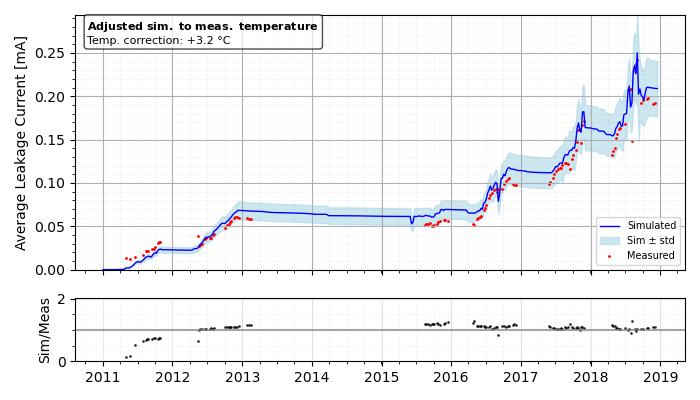}
}
\caption{Evolution of simulated and measured average $I_{\text{leak}}$ with a 3.2~\textdegree C temperature correction. The bottom plot shows the ratio between the simulation and measurements }
\label{fig:temp_correction}
\end{figure}

The strong dependence of leakage current on temperature has a significant impact on the results. Two NTC thermistors are placed on each hybrid: NTC00 closer to the cooling devices and NTC01 next to the module edge (see Fig.~\ref{fig:velo_module}).  Their readings within a single sensor differed up to several degrees; the temperature gradient across the sensor is therefore non-negligible and certainly affects the measured leakage current.
NTC01 readings were used in the following analysis, as closer to the expected silicon temperature.
The heat distribution on the module during collisions (when the chips are powered) is non-uniform; it varies with distance from the IP and sensor radius, due to both local heating from electronics and energy deposition by particles traversing the silicon. The temperature gradient increases the measurement uncertainties, and the recorded values are systematically underestimated. The relationship between thermistor reading and the actual silicon temperature was studied in a dedicated vacuum tank test, where the difference between $T_{\text{NTC}}$ and the (higher) $T_{\text{Si}}$ was found to be 3.2$\pm$1.6~\textdegree{}C~\cite{Casse:2008zza}. This is an average offset determined for all sensors during laboratory tests conducted before irradiation. The temperature difference was not determined individually for each sensor, and no similar measurements were performed after irradiation. Therefore, the 3.2~\textdegree C correction was adopted as the most suitable available measurement for correcting the temperature data (see Fig.~\ref{fig:temp_correction}).
%
Both the normalisation on the plots and the simulations were performed using these adjusted values. The agreement between simulation and measurement improved, and the ratio between them flattened. The mean percentage difference between the simulated and measured values changed from $-21.20\%\pm5.95$ to $9.41\%\pm8.53$ percentage points.

From a data analysis point of view, this inconsistency can be interpreted either as an overestimated current relative to the measured temperature or as an underestimated temperature relative to the observed current. Both interpretations were examined. In the first test, discussed above, for each sensor, the measured current was treated as the baseline, while the recorded temperature was considered as underestimated relative to the actual silicon temperature. In the second test, the measured temperature was assumed as the baseline, and the measured currents were scaled down. The results were nearly identical for both temperature correction approaches, and this strong agreement suggests that the overall method is consistent, the algorithm performs as intended, and the treatment of temperature appears consistent with both the Hamburg-model framework and the scaling procedures.

The results indicate that a significant fraction of the initial discrepancy between simulation and measurement indeed originates from an underestimation of the measured temperature.
Radiation damage, which increases module power consumption, may cause the needed temperature correction to change over time and differ between sensors, making the use of a single correction value a simplification. The observed disagreement in the relative currents induced in sensors at different positions may be partly attributed to this effect. However, verifying this hypothesis is difficult without data on changes in the temperature distribution across the sensor resulting from irradiation. Nevertheless, this approach proved useful and significantly improved the agreement.

For all subsequent plots, the correction was applied as follows: all temperature measurements were increased by a constant value of 3.2~\textdegree C. These corrected values consistently replaced the original measurements and were used as the annealing temperature in the simulation and as the basis for applying the temperature correction (Equation~\ref{eq:R_for_i}).
\subsection{Comparison of leakage-current evolution across sensors}
\label{sec:across_sens}
\begin{figure}
\centering
\resizebox{0.49\textwidth}{!}{%
  \includegraphics{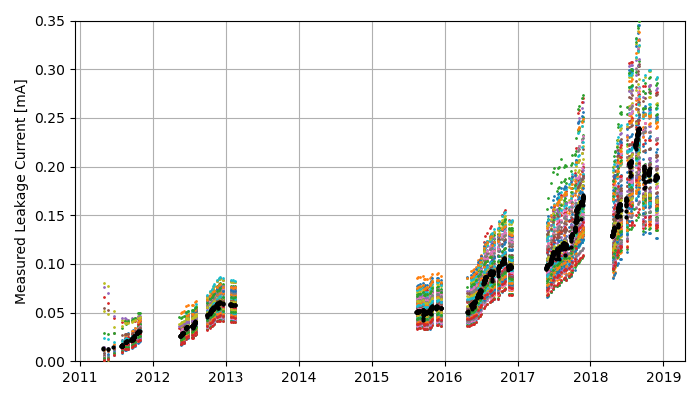}
}
\caption{Measured leakage current of VELO sensors. Black points represent the average values. }
\label{fig:meas}
\end{figure}

\begin{figure}
\centering
\resizebox{0.49\textwidth}{!}{%
  \includegraphics{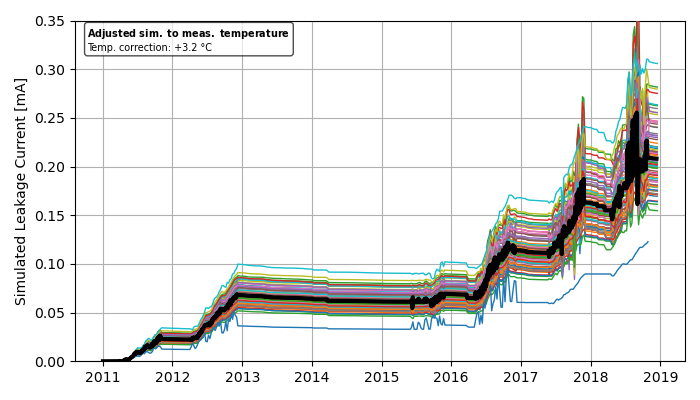}
  }
\caption{Simulated leakage current of VELO sensors.  The NTC temperature readings used in the simulation include a correction of +3.2~\textdegree C. The bold black line represents the average value. }
\label{fig:sim}
\end{figure}

Fig.~\ref{fig:meas} and~\ref{fig:sim} show the current in time evolutions of $I_{\text{leak}}$ for the VELO sensors, with the former showing the measured data and the latter the corresponding simulations. 
In the data from the early operation period of VELO, around 2011, an initial surface-generated current is visible~-~it originates from the production phase, differs between sensors, and gradually decreases under irradiation. After delivering approximately 2.5\,fb$^{-1}$ of integrated luminosity, its contribution becomes negligible. The increasing spread of the curves is an expected phenomenon: the particle flux is nonuniform and kinetic energy is absorbed along the trajectory, so the fluence received by the sensors varies significantly with $z$-position. 

\begin{figure}
\centering
\resizebox{0.48\textwidth}{!}{%
  \includegraphics{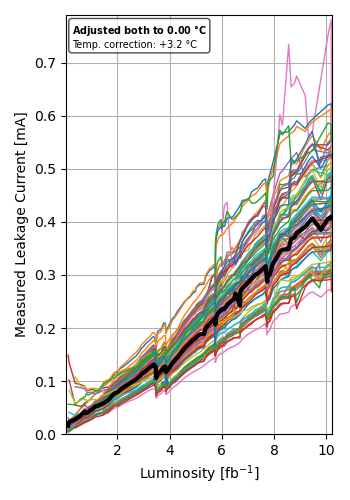}
  \includegraphics{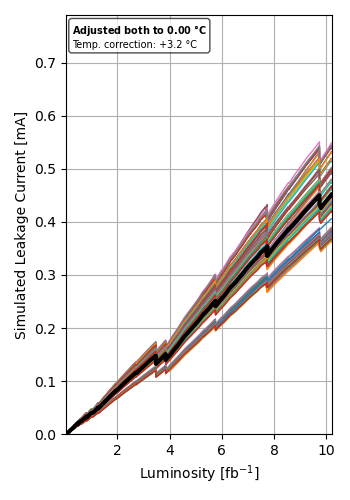}
}
\caption{Evolution of $I_{\text{leak}}$ for all sensors as a function of integrated luminosity. The left-hand panel shows measured currents and the right-hand panel simulated currents, both normalised to 0~\textdegree C. The bold black line represents the average value. Sudden jumps appear on the plot during periods without irradiation, resulting from annealing.}
\label{fig:i_vs_lumi}
\end{figure}

\begin{figure}
\centering
\resizebox{0.48\textwidth}{!}{%
  \includegraphics{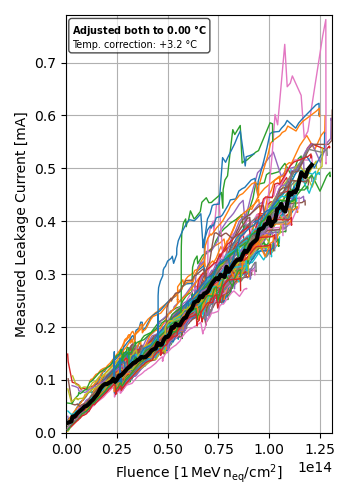}
  \includegraphics{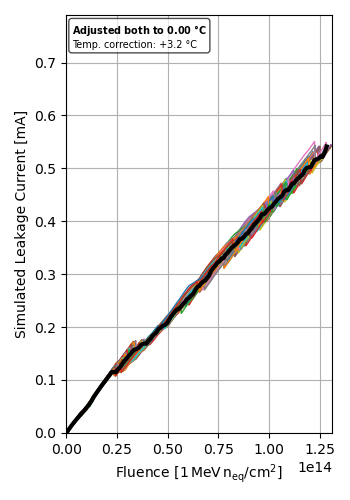}
}
\caption{Evolution of $I_{\text{leak}}$ for all sensors as a function of fluence. The left-hand panel shows measured currents and the right-hand panel simulated currents, both normalised to 0~\textdegree C. The bold black line represents the average value. Lines corresponding to sensors that received different levels of irradiation appear uncorrelated, because the sensors were irradiated at different rates. 
}
\label{fig:i_vs_fluence}
\end{figure}

Some outlier sensors are noticeable, likely due to inaccuracies in the measured temperature or possible sensor breakdowns, for example, the dark-blue curve lying significantly below the others in the time-dependent simulated data was identified. This result is further observed in the evolutions plotted with respect to delivered luminosity (Fig.~\ref{fig:i_vs_lumi}) and fluence (Fig.~\ref{fig:i_vs_fluence}), where the currents were normalised to 0~\textdegree C, in order to obtain a temperature-independent trend. At least some of the observed anomalies appear to be time-correlated. The linear dependence of the current-related damage on $\Phi_{eq}$ is reproduced in the simulation and observed in the measurements when the temperature scaling based on measured values is applied. This suggests that the observed jumps are associated with faulty temperature readings. The identified outlier sensors (13, 15, 40, 70, 79, 95 and 101) were therefore excluded from the plots, mean values and other quantitative analyses presented in the paper. Their exclusion had only a minor impact on the results.

\begin{figure}
\resizebox{0.49\textwidth}{!}{%
  \includegraphics{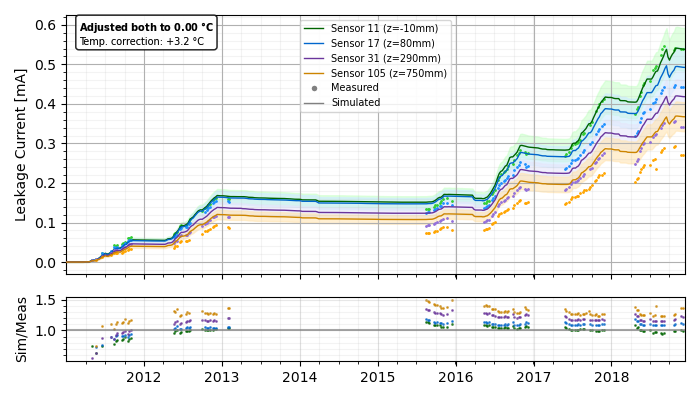}}
\caption{Leakage current evolution comparison for sensors at increasing distances from the IP. The bottom plot represents the ratio between simulation and measurement. Both sets of current data were normalised to 0~\textdegree C, and a temperature correction was applied. }
\label{fig:4_sensors}      
\end{figure}

Leakage current growth is more significant for sensors located closest to the collision point, which is obvious. However, the cause of much wider spread across sensors than expected (see Fig.~\ref{fig:i_vs_lumi} and \ref{fig:i_vs_fluence}) remains unclear; the simulation shows a weaker dependence on sensor distance from the IP than is observed in the measurements, see Fig.~\ref{fig:4_sensors}. Sensors 31 and 105, located further away from the IP, are overestimated by the simulation by more than 20\% in most cases. The discrepancy becomes less pronounced when moving closer to the interaction point, up to sensor 11, which started to be underestimated towards the end of 2017. In terms of time, the largest discrepancy in all cases occurs at the beginning of Run~2. However, this is expected to be largely due to uncertainties related to the prolonged annealing period during LS1, as the available records of the detector storage conditions contain gaps in time.
%
\subsection{Fluence dependence on the relative sensor position}
\begin{figure}
\centering
\resizebox{0.47\textwidth}{!}{%
  \includegraphics{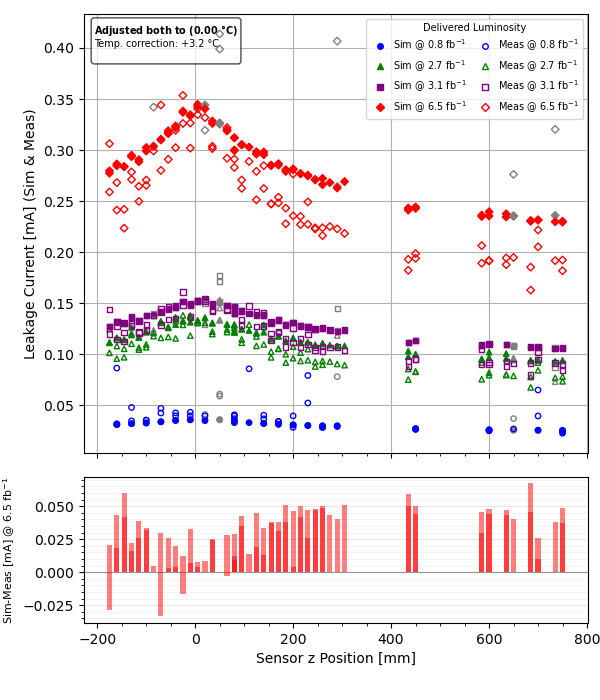}}
\caption{Leakage current as a function of sensor position, normalised to 0~\textdegree C, showing positional trends. The bottom panel shows the difference between simulated and measured currents. Sensors previously identified as anomalous are marked in grey, and the corresponding differences are not shown. }
\label{fig:sensor_pos_temp}      
\end{figure}

To help illustrate this issue globally, Fig.~\ref{fig:sensor_pos_temp} shows the leakage current of each VELO sensor as a function of its position along the beam axis at four values of integrated luminosity. In order to reveal the spatial dependence of the current, temperature normalisation was applied. The data correspond to time points when the integrated luminosity reached approximately 0.8, 2.7, 3.1, and 6.5~fb$^{-1}$, in order to compare it with measurement results based on IT scans data \cite{Akiba:2018rhn} (results were also cross-checked with~\cite{Buchanan:2018jwc}). However, the cited publications do not provide simulated values or account for temperature corrections. Therefore, a quantitative comparison including these additional aspects of the present analysis is not possible.

Most particles are produced at low polar angles, resulting in a higher particle flux traversing the edges of the sensors located further downstream~\cite{OblakowskaMucha2018}, which compensated for part of the distance-related decrease. This is reflected in the curve flattening as $z$ increases. 
A noticeable difference in the shape of the simulated and measured data is observed. The simulated evolutions are more compact, while the measured ones exhibit greater differentiation between the least and most irradiated sensors. 
The mismatch appears consistent with ATLAS pixel detector (IBL) fluence-to-luminosity conversion factor studies~\cite{Lari:2020wzo}. 

\begin{figure}
\centering
\resizebox{0.47\textwidth}{!}{%
  \includegraphics{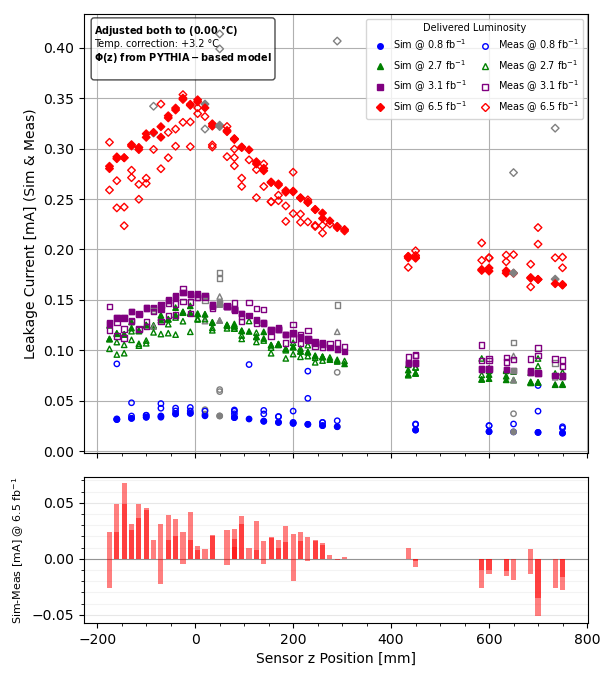}}
\caption{Same as Fig.~\ref{fig:sensor_pos_temp}, but using an alternative model with a different relative $\Phi(z)$ dependence.}
\label{fig:sensor_pos_temp_pythia}      
\end{figure}

To investigate the impact of the spatial fluence distribution received from FLUKA on the results, the leakage current simulation was repeated using a rescaled fluence distribution based on the \(\Phi(z)\) dependence obtained in an independent analysis of the VELO~\cite{Hickling:2011mxa}. The maximum fluence value was kept unchanged and used as a reference for rescaling the remaining values to reproduce the spatial distribution obtained with the referenced PYTHIA-based model. Consequently, the results should be interpreted in terms of the relative leakage currents of the sensors rather than their absolute values. See Fig.~\ref{fig:sensor_pos_temp_pythia}, 
which appears to reproduce the relative fluence among sensors located at different distances from the interaction point considerably better. The improvement is more pronounced for sensors at positive $z$-positions, whereas the agreement for sensors at negative $z$-positions remains less satisfactory. This is effect of the IP location uncertainty, which has lead to a different sensor being identified as the most highly irradiated, affecting the rescaling procedure. Consequently, the spatial distribution is shifted towards negative $z$-values.

\begin{figure}
\centering
\resizebox{0.475\textwidth}{!}{%
  \includegraphics{plots/lumi_meas_adj-both-0_deli_ntc01_tcor-temp_small_EPJ_revision.png}
  \includegraphics{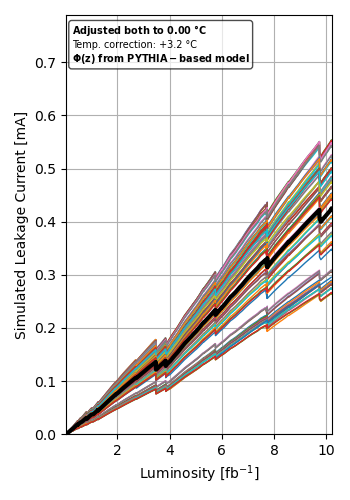}
}
\caption{Evolution of $I_{\text{leak}}$ for all sensors as a function of integrated luminosity, obtained using an alternative model with a different relative $\Phi(z)$ dependence. Compared to Fig.~\ref{fig:i_vs_fluence}, the alternative model provides a better description of the relative rates of leakage current increase among sensors at different positions along the $z$-axis.}
\label{fig:i_vs_lumi_pythia}
\end{figure}

The discrepancy in the spread of leakage currents among sensors at different positions along the \(z\)-axis was smaller (see Fig.~\ref{fig:i_vs_lumi_pythia}), suggesting a more accurate prediction of the relative rates of leakage current increase across the sensors. However, other sources of uncertainty dependant on sensor position, such as the temperature gradient, may also affect the spatial pattern of the relative leakage currents.

Obtaining a reliable estimate of $\Phi$ is highly dependent on the model properties, especially on particle production, secondary processes and radiation-material interaction. This leads to significant differences between the generators, including substantial differences in the fluence~\cite{Oblakowska-Mucha:2020jjt}. However, the conclusions are non-trivial and the sources of these discrepancies is difficult to identify at this point; the large number of variables involved implies that more detailed studies based on larger datasets are needed.

While analysing these past studies might be challenging nowadays, and therefore performing a quantitative and precise analysis is difficult, the model comparison presented here highlights the importance of the model choice and demonstrates that future analyses must treat both the modelling assumptions and the interpretation of the results with particular caution.
\section{Conclusion}
The leakage-current evolution model was successfully implemented for the VELO sensors, with a new software enabling extensive testing. Measurements of the leakage current were compared with predictions from the Hamburg model simulation. The integrated luminosity and temperature data were taken from actual measurements, improving precision compared to prediction-based studies. The resulting plots display the increase of leakage current with fluence, the annealing behaviour of the damage-related parameter $\alpha$, temperature fluctuations, and specific periods in the detector’s history, such as a controlled increase in temperature to induce annealing at the end of 2018. 

Multiple tests were performed, including those involving sensor groups, their distances from the IP and temperature normalisation procedures. Many aspects of the analysis were approached from different perspectives, increasing the overall reliability of the findings and reducing the risk of concluding results potentially biased by the simulation software. 

Overall, good agreement was found between the collected and predicted data, especially considering the complexity of both the simulation and measurement processes. Potential sources of uncertainty include inadequate temperature readings (tested in presented study), the Hamburg model parameters, and the FLUKA fluence simulation including the NIEL-scaling factors~\cite{VasilescuLindstroem2025}. Some of these uncertainties may be substantial. According to recent discussions~\cite{Dawson:2021waf,huhtinen2019uncertainties}, the uncertainties associated with the NIEL-scaling factors might be as high as 30\%. Since those serve as input to the simulation software, they can adversely affect the results. The Hamburg model parameters are difficult to test and validate reliably, even with access to leakage current evolution data from multiple sensors irradiated at different rates. The large number of available parameters provides considerable flexibility in tuning the model, allowing the parameters to be adjusted to achieve an almost perfect agreement between simulated and measured current evolution. However, given the other sources of uncertainty discussed above, many of which may also exhibit substantial variability, such agreement would not provide trustworthy parameter estimates or reflect the actual physical processes. A better understanding of these factors and constants, as well as improved possibilities for testing them, may become available in the future, for example, once discrepancies in temperature measurements are better understood or addressed through more accurate measurements. Nevertheless, the potential uncertainties introduced by the Hamburg model itself shouldn't be forgotten.

It is also worth mentioning that according to recent studies $E_{\mathrm{eff}}$ may depend on fluence; first results indicate that it decreases after heavy irradiation~\cite{OblakowskaMucha2018}, although the available data 
are still too limited to include this effect reliably at the present stage, especially since they are based on IT scans, which are inherently subject to temperature measurement bias.

The systematic underestimation of the sensor temperature appears because it was measured by devices located relatively far from the silicon volume, and heat distribution across the hybrid is non-uniform. This is particularly problematic due to the strong dependence of leakage current on temperature, affecting both the normalisation of the measured current and the simulation, as the NTC measurements serve as input to both. In this study, a constant temperature correction, measured before detector irradiation, has been applied. More precise temperature corrections accounting for variations in the temperature gradient after irradiation and as a function of sensor position have not been measured. This remains a limitation of the precision of the method.

Preliminary investigations of this issue yielded results consistent with the underlying assumptions. While this topic has been omitted in most publicly presented studies, considering the growing challenges for future silicon sensors, taking it into account and conducting an in-depth investigation may be crucial for obtaining reliable predictions for the operation of next-generation silicon detectors.

A weaker-than-expected dependence of leakage current on sensor position was observed, and similar discrepancies have been reported in independent studies from the ATLAS experiment. This may originate from the FLUKA (DPMJET) simulation used at an earlier stage of the analysis. Further studies are required to compare different models.

Improved precision, a more detailed discussion of input parameters and their uncertainties, and the inclusion of additional data may enable a more robust evaluation of the model parameters and support further studies on radiation-induced damage.
%
%
%

%
\bibliographystyle{unsrturl}
\bibliography{references}

\end{document}